\documentclass[11pt,a4paper]{article}
\usepackage{eta_etap_style}
\newcommand{\comment}[1]{}
\newcommand{\beq}{\begin{equation}}
\newcommand{\eeq}{\end{equation}}
\newcommand{\bea}{\begin{eqnarray}}
\newcommand{\eea}{\end{eqnarray}}

\newcommand{\gsim}{\lower.7ex\hbox{$
\;\stackrel{\textstyle>}{\sim}\;$}}
\newcommand{\lsim}{\lower.7ex\hbox{$
\;\stackrel{\textstyle<}{\sim}\;$}}

\newcommand{\etp}{\eta ^{\prime}}

\newcommand{\bet}{B \rightarrow K\, \eta}

\newcommand{\eod}{\end{document}}

\def\cp{{\bf CP}}

\newcommand{\SU}{SU(3)_{fl}}

\title{$\boldmath{
\eta - \eta^{\prime}}$ Mixing -- From electromagnetic transitions to weak decays of
charm and beauty hadrons
}

\author[a]{C. Di Donato,}
\author[b,a]{G. Ricciardi,}
\author[c]{I.I. Bigi}

\affiliation[a]{I.N.F.N. Sezione di Napoli,\\
Complesso Universitario di Monte Sant'Angelo --- via Cintia, 80126 Napoli, Italy}
\affiliation[b]{Dipartimento di Scienze Fisiche, Universit\'a di Napoli Federico II,\\
Complesso Universitario di Monte Sant'Angelo --- via Cintia, 80126 Napoli, Italy}
\affiliation[c]{Department of Physics, University of Notre Dame du Lac,\\
Notre Dame, IN 46556, USA}

\emailAdd{camilla.didonato@na.infn.it}
\emailAdd{giulia.ricciardi@na.infn.it}
\emailAdd{ibigi@nd.edu}

\abstract{
It has been realized for a long time that knowing the $\eta$ and $\eta^{\prime}$ wave functions in terms of
quark and gluon components probes our understanding of nonperturbative QCD dynamics. Great
effort has been given to this challenge, yet no clear picture has emerged even with the
most recent KLOE data. We point out which measurements would be most helpful in arriving at a
more definite conclusion. A better knowledge of these wave functions will significantly help to
disentangle the weight of different decay
subprocesses in semileptonic decays of $D^+$, $D_s^+$ and $B^+$ mesons. The resulting insights
will be instrumental in treating even nonleptonic
$B$ transitions involving $\eta$ and $\eta^{\prime}$ and their \cp~asymmetries; thus they can
sharpen the case for or against new physics intervening there.}


\begin{document}

\begin{flushright}
UND-HEP-11-BIG\hspace*{.08em}03\\
DSF-6-2011
\end{flushright}

\maketitle
\flushbottom
\section{Introduction}

The question of $\eta - \eta^{\prime}$ mixing
\footnote{The term ``mixing" is often used when oscillations, e.g.,
$B^0 - \bar B^0$ are involved;
however with oscillations one has a nontrivial time evolution, but not for
$\eta$ and $\eta^{\prime}$ mixing.}, i.e., how their wave functions are composed of
$\SU$ singlet and octet $\bar qq$ components, goes back to the beginning of the quark model
era \cite{QMREF1,QMREF2,QMREF21, QMREF3,QMREF4,QMREF5,QMREF6,Akhoury:1987ed, QMREF7}.
With the advent of QCD it became even more involved, since QCD brought with it more dynamical
degrees of freedom, namely, gluons, which can form a second class of $\SU$ singlets. Determining
$\eta - \eta^{\prime}$ mixing is thus an intriguing element in understanding QCD's
nonperturbative dynamics. Lattice QCD's attempts to establish theoretical control over this
mixing are still in their infancy \cite{Christ:2010dd,Dudek:2011tt}.
Showing that there is a purely gluonic component in the
$\eta$ and/or $\etp$ wave functions would establish for the first time that gluons, which have been
introduced to mediate the strong interactions and whose presence as independent
degrees of freedom has been demonstrated as progenitors of jets
in `hard' collisions, play an independent role also in hadronic spectroscopy. In section \ref{EEPEVI}
we introduce basic notions relevant for $\eta-\eta^{\prime}$ mixing, while in
section \ref{PHEN} we review the somewhat ambivalent findings from several phenomenological studies.  Armed with this knowledge we discuss weak $D$ and $B$ decays producing $\eta$ and $\etp$
mesons in section \ref{WEAKDBDEC} and what the
observed rates can tell us about the underlying quark level transitions; we comment briefly on
how the structure of the  $\eta$ and $\etp$ wave functions affect \cp~asymmetries in the
channels $B_d \to \etp K_S$ and  $B_d \to \eta K_S$. Finally in
section \ref{OUT} we present a summary and outlook.

\section{$\eta-\eta^{\prime}$ Mixing
\label{EEPEVI}}

Based on approximate QCD flavor $SU(3)_{fl}$ symmetry, the mixing of the $\eta$ and $\eta^{\prime}$ mesons can be described in two different bases:
\begin{enumerate}
\item
the $SU(3)_{fl}$ singlet and octet components
$|\eta_0\rangle = \frac{1}{\sqrt{3}}|u \bar u + d \bar d + s \bar s\rangle$ and
$|\eta_8\rangle=\frac{1}{\sqrt{6}} |u \bar u + d \bar d - 2 s \bar s\rangle$,
respectively:
\begin{eqnarray}
\left (
\begin{array}{c}
|\eta \rangle \\
|\eta^{\prime} \rangle \\
\end{array}
\right ) =
\left (
\begin{array}{cc}
\cos \theta_P & -\sin \theta_P \\
\sin \theta_P & \cos \theta_P \\
\end{array}
\right )
\left (
\begin{array}{c}
|\eta_8 \rangle \\
|\eta_0 \rangle \\
\end{array}
\right )  \; ;
\label{mixsu3}
\end{eqnarray}
\item
the quark-flavor basis with
$|\eta_q\rangle =\frac{1}{\sqrt 2}|u \bar u + d \bar d\rangle $ and
$|\eta_{s} \rangle = |s \bar s\rangle $:
\begin{eqnarray}
\left (
\begin{array}{c}
|\eta \rangle \\
|\eta^{\prime} \rangle \\
\end{array}
\right ) =
\left (
\begin{array}{cc}
\cos \phi_P & -\sin \phi_P \\
\sin \phi_P & \cos \phi_P \\
\end{array}
\right )
\left (
\begin{array}{c}
|\eta_q \rangle \\
|\eta_{s} \rangle \\
\end{array}
\right )  \; .
\label{mixqf}
\end{eqnarray}
\end{enumerate}
As long as state mixing is regarded,
one may freely transform from one  basis to the other;  the two parametrizations  are related through
\begin{equation}
\theta_P = \phi_P - \arctan \sqrt 2 \simeq \phi_P - 54.7 ^{\circ}
\label{relmix}
\end{equation}
In the $SU(3)_{fl}$  symmetry limit, $\theta_P=0$, and $\phi_P$ takes the so-called `ideal'  value $\phi_P= \arctan \sqrt 2  \simeq 54.7 ^{\circ}$.

Just for orientation: the quadratic [linear] Gell-Mann Okubo (GMO) mass formula points to $\theta_P\simeq -10^{\circ}$,
$\phi_P\simeq 44.7^{\circ}$
[$\theta_P\simeq -23^{\circ}$, $\phi_P\simeq 31.7^{\circ}$].

The mixing schemes have been
analyzed in the context  of chiral perturbation theory.
On lattice, it is not an easy task to study $\eta$ and $\eta^\prime$, as experienced in the last decade of attempts.
The RBC-UKQCD Collaboration has reported a pioneering calculation of the $\eta$ and
$\eta^\prime$
 masses and mixing angle of $\theta_P = -14.1(2.8)^\circ$
 using $N_f = 2+1$ flavor domain wall ensembles on an Iwasaki gauge action \cite{Christ:2010dd}.
 Their results show small octet-singlet mixing, consistent with the quadratic GMO within the large statistical errors. Masses and mixing angle of the $\eta$ and
$\eta^\prime$ have also been calculated by the Hadron Spectrum Collaboration \cite{Dudek:2011tt}, using lattice QCD with unphysically heavy light (up, down)
quarks and  a single lattice spacing: their estimate value is $\phi_P = 42(1)^{\circ}$. The  large value of the mixing angle $\phi_P$ in the pseudoscalar sector,  with respect to  other ones (e.g.  the vector mesons $|\omega> \simeq |\eta_q> $ and  $\phi\simeq |\eta_s> $,  with a  mixing angle   $\phi_V= (3.4 \pm 0.2)^\circ$ \cite{Bramon:2000fr}) is expected, because of the additional mixing induced  by the axial $U(1)$ anomaly (\cite{Feld} and references therein).


In the 1990s the possibility of a single
angle description being inadequate started to be considered.
Several papers \cite{Schechter:1992iz, Ball:1995zv, Leut1,Leut2,Feld&K,Feld&S,Feld, Feldmann:2002kz, Escribano:2005qq}, based on theoretical studies as well as  on comparison with data,  pointed out that
 the pattern of $SU(3)_{fl}$ breaking requires a  description  in terms of two angles.
Phenomenological analyses have often involved weak
decay constants  $f_{\eta^{(\prime)}}^a$,  defined by the relation
 $ < 0| A^a_\mu | \eta^{(\prime)}(p) > = i f_{\eta^{(\prime)}}^a \, p_\mu$.
In the octet-singlet basis $a=8,0$ and $ A_\mu^{8,0}$ are the octet and singlet axial-vector currents;
in the quark-flavor basis, $a=q,s$ and $ A_\mu^{q,s}$ are the nonstrange and strange axial-vector currents.
Because of $SU(3)_{fl}$ breaking, the mixing of the decay constants does not necessarily follow the same pattern as the state mixing (see e.g. \cite{Feldmann:2002kz, Escribano:2005qq}).
For completeness, we report here the most general parametrizations involving two independent axial-vector currents
and two different physical states:
\begin{enumerate}
\item
\begin{eqnarray}
\left (
\begin{array}{cc}
f^8_{\eta} & f^0_{\eta} \\
f^8_{\eta^{\prime}} & f^0_{\eta^{\prime}} \\
\end{array}
\right ) =
\left (
\begin{array}{cc}
f_8 \cos \theta_8 & -f_0 \sin \theta_0 \\
f_8 \sin \theta_8 & f_0 \cos \theta_0 \\
\end{array}
\right )
\label{mixsu3dec}
\end{eqnarray}
\item
\begin{eqnarray}
\left (
\begin{array}{cc}
f^q_{\eta} & f^s_{\eta} \\
f^q_{\eta^{\prime}} & f^s_{\eta^{\prime}} \\
\end{array}
\right ) =
\left (
\begin{array}{cc}
f_q \cos \phi_q & -f_s \sin \phi_s \\
f_q \sin \phi_q & f_s \cos \phi_s \\
\end{array}
\right )
\label{mixqfdec}
\end{eqnarray}
\end{enumerate}
We observe that in Eq. (\ref{mixsu3dec}) as in Eq. (\ref{mixsu3}) the angles are chosen in such a way that $\theta_P= \theta_8=\theta_0=0$  corresponds to the $SU(3)_{fl}$
symmetric world.
As before any expression in one scheme can be translated into the other one in a straightforward mathematical way.
However different dynamical implementations of $SU(3)_{fl}$ breaking
suggest a different ansatz; for example it has been suggested that attributing $SU(3)_{fl}$ breaking
to Okubo-Zweig-Iizuka (OZI) violating contributions leads to $\phi_q \simeq \phi_s$, recovering a description in terms
effectively of a single angle in the quark-flavor basis \cite{Feld, Leut2}. As it is well known, the OZI rule leads to a
suppression of strong interaction processes where the final states can only be reached through quark anti-quark
annihilation.
In the octet-singlet basis,  instead, the differences in $\theta$ may be sizable, and most analyses find the range
$\theta_8-\theta_0 \approx [-19^\circ, -12^\circ]$ (\cite{Ball:1995zv, Leut1,Leut2,Feld, Escribano:2005qq} and references therein).
In this respect, the quark-flavor basis plays a privileged role; we will use such a basis
in the following, assuming a single mixing angle $\phi_P = \phi_q = \phi_s$, that correspond to Eq. (\ref{mixqf}).
We can see from Eq. (\ref{mixqfdec}) that under this assumption the decay constants follow the same pattern of particle state mixing.

The plot thickens further still in QCD, for one can form an $SU(3)_{fl}$ singlet not only from quark-antiquark combinations,
but also from pure gluon configurations with the simplest one being
a $gg$ combination. Since in general all components compatible
with the quantum numbers of a state can appear in that state's wave function, there is no {\em a priori}
reason why the $\eta$ and $\eta^{\prime}$ wave functions could not contain
such configurations. On general grounds they will contain also $c \bar c$ (or $b \bar b$) components,
but probably on a significantly smaller level, since the
mass scale for gluonic excitations is presumably lower than the $J/\psi$ mass; therefore we
will ignore $c \bar c$ (and $b \bar b$) admixtures in our subsequent analysis.
Using the quark-flavor basis,  we write down \cite{QMREF21}:
\begin{eqnarray}
|\eta^{\prime}\rangle &\simeq & X_{\eta^{\prime}}|\eta_q \rangle +
Y_{\eta^{\prime}}|\eta_s \rangle + Z_{\eta^{\prime}}|gg \rangle \nonumber \\
|\eta\rangle &\simeq & X_{\eta}|\eta_q\rangle + Y_{\eta}|\eta_s \rangle + Z_{\eta}|gg \rangle
\label{colla}
\end{eqnarray}
One would  expect the heavier $\eta^{\prime}$ to contain a higher dose of gluonic
components than the $\eta$, which is also mainly an $SU(3)_{fl}$ octet.
Setting $ Z_{\eta}$ to zero is presumably a pragmatically sound approximation.
In \cite{Li} the authors use a number of  parameterization schemes to analyze
 $J/\psi$ and $\psi^\prime$ decays into vector and pseudoscalar mesons; in most cases they find a value for the gluonic content of $\eta$ compatible with zero, with an exact numeric value of $Z_\eta^2/Z_{\eta^\prime}^2$ that is strongly model dependent  and ranges from $  10^{-11}$ to $ 0.08$. Reference \cite{Li} also presents a framework, based on old perturbation theory, that allows a much higher gluonic content in
$\eta$, that is $Z_\eta^2/Z_{\eta^\prime}^2 \approx 1$. This result is inconsistent with the analysis of the same data made in \cite{Thomas},
where $ Z_{\eta}=0$ is assumed.

In the following, we use the approximation $Z_\eta=0$, $ Z_{\eta^{\prime}} \neq 0$ and we  parameterize the two orthonormal states in terms
of $\phi_P$ plus an additional mixing angle $\phi_G$:
\begin{eqnarray}
|\eta^{\prime}\rangle &\simeq & \cos  \phi_G \sin  \phi_P |\eta_q \rangle +\cos  \phi_G
\cos \phi_P|\eta_s \rangle + \sin \phi_G |gg \rangle \nonumber \\
|\eta\rangle &\simeq & \cos   \phi_P |\eta_q\rangle - \sin  \phi_P |\eta_s \rangle
\label{colla1}
\end{eqnarray}

As already mentioned it is unlikely that lattice simulations of QCD will determine the $\eta$
and $\eta^{\prime}$ wave functions in the {\em near} future.
Phenomenological studies are thus our only recourse.  Several such analyses have been undertaken recently: while their findings are not inconsistent, their messages are ambivalent, as we will discuss in the next section.

\section{Phenomenological Studies of $\eta-\eta^{\prime}$ Mixing}
\label{PHEN}

There are three classes of electromagnetic and strong transitions that can provide information on the mixing angles and the gluonic content:
\begin{itemize}
\item
Radiative vector and pseudoscalar meson decays:
\bea
& & \psi^\prime, \, \psi , \,  \phi  \to \gamma \eta^{\prime} \; \; vs. \; \;  \gamma \eta \nonumber \\
& &  \rho, \, \omega  \to \gamma \eta  \nonumber \\
& &  \eta^{\prime} \to \gamma    \omega,  \, \gamma \rho
\label{PVbelow}
\eea
\item
Decays into two photons or production in $\gamma\gamma$ collisions:
\bea
\eta^{\prime} \to \gamma  \gamma \; \; &vs.& \; \; \eta \to  \gamma \gamma \label{PGG} \\
\gamma \gamma \to  \eta \; \; &vs.& \; \; \gamma \gamma \to  \eta^{\prime} \label{fusion}
\eea
\item
Decays of $\psi$ into PV final states with the vector meson acting as a `flavor filter':
\beq
\psi \to \rho /\omega /\phi + \eta \; \; vs. \; \; \eta^{\prime}
\label{PVabove}
\eeq

\end{itemize}

\subsection{Present Status}
\label{PRESTAT}
Recent papers on the glue content of the $\eta^{\prime}$ by KLOE \cite{kloe07} and Li et al. \cite{Li}
have motivated other studies of a range of
different processes \cite{Rafel,Thomas,Ambrosino:2009sc}.
Escribano, Nadal \cite{Rafel, RafelJPsi}  and Thomas \cite{Thomas} have analyzed all processes of Eq. (\ref{PVbelow}), (\ref{PGG}),
(\ref{fusion}) and (\ref{PVabove}).
The old and new analysis from KLOE \cite{kloe07,Ambrosino:2009sc} and Escribano and Nadal \cite{Rafel} refer
to those processes of Eq. (\ref{PVbelow}) whose dynamical scale is below 1.02 GeV (that is, including
$\phi$, but excluding $\psi$ and $\psi ^{\prime}$ decays), while Li et al. \cite{Li} have analyzed the ones above 1.02 GeV.
In Table \ref{phidectab1} we have summarized the results of \cite{Rafel,Thomas,kloe07},
based on radiative decays of vector/pseudoscalar mesons below 1.02 GeV.

The KLOE analysis also includes constraints from $\pi^0/\eta^{\prime} \to  \gamma \gamma$, according to the
prescription of Ref. \cite{Kou:1999tt}.
These results are obtained  by including vector-pseudoscalar wave function overlaps, assuming the $\eta^{(\prime)}$ to be
a pure $q \bar q$ state, i.e. $Z^2_{\eta^{(\prime)}}=0$, and the dependence of the decay widths on the mixing angle as
in \cite{Rafel}.

\begin{table}[t]
\centering
\vskip 0.1 in
\begin{tabular}{|l|c|} \hline

  Analysis     &    $\phi_P$ (Ansatz $Z^2_{\eta^{\prime}}\equiv 0)$\\
\hline
 KLOE  & $(41.3\pm0.3_{stat}\pm0.7_{sys})^{\circ}$ \\
 Escribano I &  $(41.5\pm1.2)^{\circ}$ \\
 Escribano II &  $(42.7\pm0.7)^{\circ}$ \\
 Thomas I  &  $(41.3\pm0.8)^{\circ}$ \\
 Thomas II   &  $(41.7\pm0.5)^{\circ}$ \\
 Thomas I with form factors & $(41.9\pm1.1)^{\circ}$ \\
 Thomas II with form factors  & $(42.8\pm0.8)^{\circ}$\\
\hline
\end{tabular}
\caption{ \it Fit values for the $\eta-\eta^{\prime}$ mixing angle as inferred by different authors from radiative decays  of vector/pseudoscalar mesons below 1.02 GeV,  assuming $Z^2_{\eta^{\prime}}=0$. Only the KLOE analysis includes also constraints from $\eta^{\prime} \to \gamma \gamma$. I labels the results from the analysis without including the latest data
on $\phi \to \eta^{\prime} \gamma$ (KLOE) and $(\rho, \omega, \phi) \to \eta \gamma$
(SND), while II indicates the same analyses performed including them.}
\label{phidectab1}
\end{table}
We see that the different analyses yield very consistent values for the mixing angle, namely
 $\phi_P \simeq 42^{\circ}$, which happens to be close to the value suggested by the {\em quadratic}
 GMO mass formula. Including the latest data from $KLOE$ \cite{kloe07} and $SND$ \cite{SND06}
 does not cause a significant shift.

In Table \ref{phidectab2} results from the same studies are listed, now allowing for a gluonic
component in $\eta^{\prime}$, i.e. $Z^2_{\eta^{\prime}} \neq 0$.

\begin{table}[t]
\centering
\vskip 0.1 in
\begin{tabular}{|l|cc|} \hline
 Analysis      &    $\phi_P$ & $Z^2_{\eta^{\prime}}$\\
\hline
 KLOE  & $(39.7\pm0.7)^{\circ}$ & $0.14\pm0.04$\\
 Escribano I &  $(41.4\pm1.3)^{\circ}$ & $0.04\pm0.09$\\
 Escribano II &  $(42.6\pm1.1)^{\circ}$ & $0.01\pm0.07$\\
 Thomas I  &  $(41.3\pm0.9)^{\circ}$ & $0.04\pm0.06$\\
 Thomas II   &  $(41.7\pm0.5)^{\circ}$ & $0.04\pm0.04$\\
 Thomas I with form factors & $(41.9\pm1.1)^{\circ}$ & $0.10\pm0.06$\\
 Thomas II with form factors  & $(41.9\pm0.7)^{\circ}$ & $0.10\pm0.04$\\
\hline
\end{tabular}
\caption{ \it Fits allowing for a gluonium component using radiative decays of vector/pseudoscalar mesons below 1.02 GeV. Only KLOE analysis includes also constraints from $\eta^{\prime} \to \gamma \gamma$.
I again labels the results from analyses without including the latest data
on $\phi \to \eta^{\prime} \gamma$ (KLOE) and $(\rho, \omega, \phi) \to \eta \gamma$
(SND), while II indicates the same analysis performed including them.}
\label{phidectab2}
\end{table}
%
The different analyses again yield consistent values for the mixing angle with
$\phi_P \simeq 42^{\circ}$ with only KLOE finding a somewhat smaller number. As
before the latest data from KLOE and SND do not cause a significant shift.
Yet while the numbers given for the size of a gluonic component are not truly inconsistent
considering the stated uncertainties, they seem to carry an ambivalent message: while the
first and last studies -- listed as ``KLOE''  and   ``Thomas with form factors'' -- point to a significant
gluonic component, the others do not. We can understand some of the differences.
As explained around Eqs. (\ref{mixsu3dec}) and (\ref{mixqfdec}) we think that assuming the mixing of the decay constants to follow the same pattern as state mixing is an oversimplification. Only Thomas
has gone beyond this assumption, and when he includes
the form factors
he finds some intriguing evidence for a gluonic contribution.

The form factors included in ``Thomas'' are phenomenological Gaussians, whose aim is  to introduce a momentum dependence for exclusive processes.
In order to understand why the findings from
``KLOE''  and ``Escribano/Thomas I-II'' for the gluonic content in Table \ref{phidectab2} are as different as they appear (for neither analysis allows for different form
factors), we can offer one comment, though:
only ``KLOE''   includes $\eta^{\prime} \to \gamma \gamma$, and that observable
pushes up the value of $Z^2_{\eta^{\prime}}$, as pointed out by Thomas.

In fact, the above theoretical discussion
has prompted the KLOE Collaboration to perform another fit \cite{Ambrosino:2009sc}, updated by
using the branching ratio values from PDG 2008 \cite{PDG08}, the more recent KLOE
results on the $\omega$ meson \cite{Ambrosino:2008gb} and using a larger number of free parameters, as suggested by \cite{Rafel,Thomas}.
The fit has been performed in the two cases: imposing the gluonium
content to be zero, that resulted in $\phi_P= (41.4\pm 0.5)^{\circ} $, or allowing it free, giving
$\phi_P=(40.4\pm 0.6)^{\circ}$.
KLOE new results confirm the  gluonium content of $\eta^\prime$
at $ 3 \sigma$ level with $Z_{\eta^\prime}^2= 0.115 \pm 0.036$, in contrast with ``Escribano/Thomas I-II'' values in Table 1.
Therefore, the actual difference between ``Escribano/Thomas I-II'' and KLOE values appears due to the inclusion in the latter of
$ \eta^\prime \rightarrow \gamma \gamma$.

The comparison presented above pointed out that decays into two photons can play a key role in the mixing parameters determination.
They can be exploited also looking at the inverse processes, namely, the production in
$\gamma \gamma$ collisions.

The L3 Collab. at LEP has published \cite{L3} the measurement of the radiative width
$\Gamma (\eta^{\prime} \to\gamma \gamma)$ produced via the collision of virtual photons,
in the reaction $e^+ e^- \to e^+ e^- \gamma^\star \gamma^\star $, $\gamma^\star \gamma^\star \to \eta^\prime $,
 $ \eta^\prime \to \pi^+ \pi^- \gamma $, using data collected at  centre-of-mass energies $ \sqrt{s} \simeq 91$ GeV.
They compare the photon-meson transition form factor with a model by Anisovich et al. \cite{Ani}, that allows a variable admixture of gluonic content, from
 $0\%$ to $15\%$. The central values of L3 data points favour a low gluonium content, but the whole interval is allowed within the large errors.

Before  L3, the same $e^+ e^- \to e^+ e^- \etp  $ reaction  had been performed at lower energy $e^+ e^-$ colliders, by using various $\eta^\prime$ decay channels (see Refs. in  \cite{L3}).
Let us review some old measurements of the radiative widths $\Gamma (\eta^{(\prime)} \to\gamma \gamma)$ used to evaluate  the mixing angles. These estimates did not consider the possibility of gluonic content and refer  to the octet-singlet basis and  the single angle approximation, whose limits   have been discussed in Sect.~\ref{EEPEVI}. To facilitate the comparison, we have quoted the results in the flavor basis, using the relation (\ref{relmix}).
The observation of $\eta$ meson production
from $\gamma\gamma$ fusion has been reported
in a 1983 Rapid Communication  by the Crystal Ball Collab.; the given mixing angle
reads $\phi_P =37.1^{\circ}\pm3.6^{\circ}$ \cite{Weinstein:1983jz}.
In 1988 they published the radiative widths for $\pi^0$, $\eta$ and
$\eta^{\prime}$ and determined mixing angles from the experimental averages, finding $\phi_P =32.3^{\circ}\pm1.2^{\circ}$ \cite{CB}.
Two years later both the MD-1 \cite{Baru:1990pc} and the ASP Collaborations \cite{ASP} presented the measurement
of the $\eta ,  \eta^{\prime} \rightarrow \gamma \gamma$ widths, with results in agreement within the errors.
The ASP Collaboration calculated the pseudoscalar mixing angle $\phi_P = 34.9^{\circ} \pm 2.2^{\circ}$ \cite{ASP}.
While these  values are compatible among them, they appear to fall significantly below those in Tables \ref{phidectab1}
and \ref{phidectab2}.

A new surge of  experimental data and updated analyses is strongly needed. The BABAR Collaboration has led the way presenting
recent studies on the $\gamma \gamma^\star \rightarrow \eta^{(\prime)}$ transition form factors in the momentum transfer range from
4 to 40 GeV$^2$ \cite{:2011hk}. They compare measured values of the $\eta^{(\prime)}$ form
factors with theoretical predictions and data for the $\pi^0$
form factor by using the description of $\eta - \etp$ mixing in the quark-flavor basis (\ref{mixqf}).
They assume no gluonic admixture and a mixing angle $\phi_P= 41^\circ$.
The  dependence on the transfer momentum of the form factor for the $|\eta_s\rangle$ state is different from
the QCD prediction \cite{Bakulev:2001pa} of the
asymptotic distribution amplitudes; data points are systematically
below the theoretical curve.
Because of the strong sensitivity of the result for the $|\eta_s\rangle$  state
to mixing parameters, an admixture of the two-gluon component
in the $\eta^{(\prime)}$ meson cannot be excluded as
a possible origin of this discrepancy.
%
%

\noindent
A new investigation is being performed by KLOE, from the analysis of off-peak data, with integrated luminosity
$L = 240$ pb$^{-1}$, already on tape, devoted to the measurement of the $\gamma\gamma \to \eta$ rate.
The off peak analysis, at $\sqrt s = 1$ GeV instead of $\sqrt s = 1.02$ GeV, allows to reduce the main background,
coming from resonant contributions $\phi \to \eta \gamma$.
After the full selection, the data set consists of 600 $\gamma\gamma \to \eta$ with $\eta \to \pi^+ \pi^- \pi^0$
and 900 $\gamma\gamma \to \eta$ with $\eta \to \pi^0 \pi^0 \pi^0$;  the cross section $\sigma(\gamma \gamma \to \eta)$
at 1 GeV is under evaluation \cite{ggetakloe}.
The upgraded KLOE detector (KLOE-2) will be  suited for taking data also at energies
away from the $\phi$ mass.
Taggers designed to detect the outcoming $e^+ e^-$ are being inserted into the KLOE detector, to
provide  a better background rejection without going off peak and allow precision measurements of the $\gamma \gamma$ cross section.
There is a proposal  to increase the DA$\Phi$NE energy up to $\sqrt s \simeq 2.5$ GeV; however, a run
at $\sqrt s \simeq 1.4$ GeV is already enough to measure the $\etp$ decay width \cite{AmelinoKloe2}.

Starting in September 2009, the Crystal Ball at MAMI has undertaken a huge upgrade, with an increase of the MAMI beam
energy and the construction and assemblage of a new tagging device; one
reason of the upgrade  is a measure of $\eta^\prime \rightarrow \gamma \gamma$ branching ratio \cite{Starostin:2011zz}.

The quoted measurements of the width are obtained with the QED process
$ e^+ e^- \rightarrow e^+ e^- \gamma^\star \gamma^\star \rightarrow e^+ e^- \eta$.
The 2010 PDG average is taken  from such experiments and gives
$ \Gamma(\eta \to \gamma \gamma) = 0.510 \pm 0.026 $ KeV. The error on the average is  5\%, while the  errors
 in individual experiments range from  8\% to 25\%.
There is a different type of measurement of $\Gamma(\eta \rightarrow \gamma \gamma)$, not included in the 2010 PDG average,
based on the Primakoff effect, where $\eta$'s are produced by the interaction of a real photon with a virtual
photon in the Coulomb field of the nucleus.
In 1974 at Cornell  a measurement
 based on the Primakoff effect gave
$\Gamma(\eta \rightarrow \gamma \gamma) =  0.324  \pm 0.046  $ KeV, a value
 $4 \sigma$ away from the QED results \cite{Browman:1974sj}. Recently, a reanalysis of the
Primakoff experiment, with a different modelling of the nuclear background,
brought the value of the width
in line with direct measurements, precisely to $ \Gamma(\eta \to \gamma \gamma) = 0.476 \pm 0.062 $ KeV
\cite{Rodrigues:2008zza}.
Extraction of the Primakoff amplitude from the data
is very delicate, because this amplitude interferes
with hadronic amplitudes due to vector meson ($\rho$ and $\omega$) and axial vector meson $b_1$ exchanges.
Increasing the energy may help, since at very high energies the growth
of the Coulomb peak must dominate over the Regge behavior of the strong amplitude.
After more than 30 years from the Cornell experiment,
a new experiment to measure the  $\Gamma(\eta \rightarrow \gamma \gamma)$ decay width via the Primakoff effect
has been proposed and  approved at  Jefferson Laboratory, using a
11.5 GeV tagged photon beam on two light targets, proton and $^4$He \cite{JLab}.
The targets have been chosen with the aim of minimizing the nuclear incoherent background and enabling a good separation
of the Primakoff production mechanism from the nuclear coherent background.
They estimate to reach a 3\% accuracy in the measurement of the $\eta$ width, that would yield less than 1$^\circ$ of
uncertainty on the $\eta-\etp$ mixing angle.

As it is well known, all $\eta$ meson possible strong decays
are forbidden in lowest order by C, CP invariance and G-parity conservation.
First order electromagnetic $\eta$ decays are forbidden as well, or occur at a suppressed rate because of involving an anomaly. The first allowed decay is therefore the second-order
electromagnetic transition $\eta \rightarrow \gamma \gamma$.
The decay $\eta \to 3 \pi$ violates isospin symmetry and it is mainly due to the isospin breaking part of
the QCD Lagrangian, since contributions from the electromagnetic interaction are strongly suppressed by chiral
symmetry \cite{sutherland}.
The main interest of this decay resides in the fact that, in principle, it offers a way to determine the mass difference of the
up-down quarks. The absolute value of the partial decay width for $\eta \to 3 \pi$ is experimentally obtained via normalization
to $\eta \rightarrow \gamma \gamma$; therefore, a change in one decay width has influence on the other \cite{PDG08}.

\begin{figure}[t]
 \centering
 \includegraphics[scale=0.35]{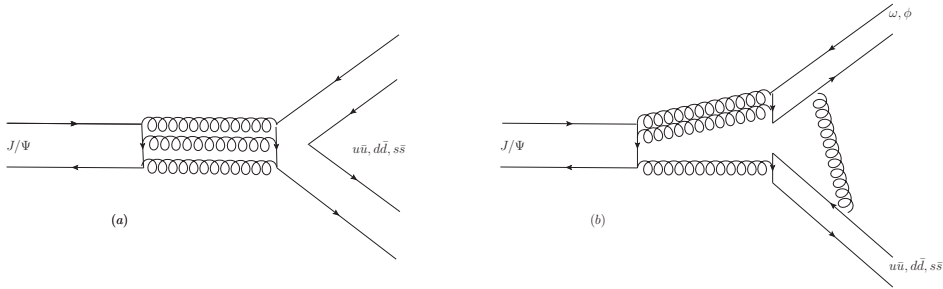}
\caption{\it (a) SOZI and (b) DOZI diagrams contributing to $\psi \to PV$ decays}
\label{fig:diag:sozidozi}
\end{figure}

A few comments are in order for the analysis of $\psi \to PV$. It was pioneered by Mark III in 1985, when they inferred from
their data $Z^2_{\eta^{\prime}}= 0.35 \pm 0.18$ \cite{MIII85}.
They assumed that such decays proceed via singly disconnected diagrams (SOZI)
with their strong quark line correlations and ignored doubly
disconnected diagrams (DOZI). In Fig. \ref{fig:diag:sozidozi}.(a) and
in Fig. \ref{fig:diag:sozidozi}.(b) we show examples of SOZI and DOZI diagrams.
Motivated by the measurement of $\psi \to \gamma\omega\phi$, which
showed the relevance of DOZI-suppressed processes in $\psi$ decays,
they performed a new analysis \cite{MIII88}, including DOZI contributions
and any additional component as gluonium or radial excitation.
The new analysis did not show evidence for non-$\bar qq$ components in the
$\eta$ and $\eta^{\prime}$ wave functions.

In 2007 Thomas \cite{Thomas} -- following the approach of Seiden et al. \cite{Seiden:1988rr}
-- investigated the strong
$\psi \to PV$ transitions;
he concluded that DOZI contributions are significant, and that any gluonium components should play a
role similar to that of DOZI contributions. From such an analysis he finds that the fit favors a small
gluonic component in the $\eta^{\prime}$, with no great significance.
Without form factors, Thomas finds $\phi_P = (45\pm4)^{\circ}$ and
$\phi_{G}=(33\pm13)^{\circ}$ (i.e. $Z^2_{\eta^{\prime}}=(0.30\pm0.21)$), whereas with form factors
$\phi_P = (46^{+4}_{-5})^{\circ}$ and  $\phi_{G}=(44\pm9)^{\circ}$, (i.e. $Z^2_{\eta^{\prime}}=(0.48\pm0.16)$).
Another phenomenological analysis of $\psi \to PV $, without form factors, by Escribano \cite{RafelJPsi},
finds $\phi_P = (40.7\pm2.3)^{\circ}$ in the hypothesis of no gluonium and, allowing for it, $\phi_P = (44.6\pm4.4)^{\circ}$
with $Z^2_{\eta^{\prime}}=(0.29^{+0.28}_{-0.26})$.

The remaining decays of the list (\ref{PVbelow}), (\ref{fusion}) and (\ref{PVabove}) are charmonium decays into $\gamma \eta^{(\prime)}$.
BESII data have better precision than previous measurements; according to the hypothesis of no gluonic contribution, SU(3) flavor
symmetry and exact OZI rule, they extract an angle in the octet-singlet scheme.
Their value, translated in the flavor scheme according to the relation (\ref{relmix}), reads $\phi_P = (32.62\pm 0.81)^{\circ}$
\cite{Ablikim:2005je}, a quite low value compared to other determinations.
The extraction of the mixing angle in \cite{Ablikim:2005je} has been performed in a very symmetric --and therefore simplified--scheme;
we observe that just by introducing a dependence on a strange/nonstrange factor, the author in \cite{Thomas}
finds for the same processes and PDG averaged data (including BESII results) values of the mixing angle in line
with determinations from other processes.
If there is any charmonium component in the $\eta^{(\prime)}$, we expect the decays of $\psi$ and $\psi^\prime$ into $\gamma \eta^{(\prime)}$
to be dominated by the magnetic dipole transition of charmonium. In that case, it is possible to estimate that the amplitudes of the
charmonium components of the $\eta^{(\prime)}$ are negligible, being less that $5 \%$ \cite{Thomas}.

\noindent
More recent  measurements of $\gamma \eta^{(\prime)}$ branching fractions have  been reported by  CLEO-c \cite{:2009tia}. The last
update of the  $\psi \rightarrow \gamma \eta^\prime$  branching fraction has been given
by BESIII \cite{Ablikim:2010kp} and reads ${\cal{B}} (\psi \to \gamma \eta^{(\prime)})
= (4.84 \pm 0.03 (stat) \pm 0.24 (sys))\, {\rm{{x}}}\, 10^{-3} $,
which is consistent with the BESII value within $1.5 \sigma$ and with the CLEO value within $1.4\sigma$.
The $\psi^\prime \rightarrow \gamma \eta^{(\prime)}$ decays have also been
observed by BESIII \cite{Ablikim:2010dx}, but no new mixing angle estimate has been reported by the Collaboration.
As far as $\Upsilon(1S) \rightarrow \gamma \eta^{(\prime)}$ is concerned, only upper limits are available for the branching
ratios from CLEO III \cite{Athar:2007hz}.

Since all extractions of the mixing angle involve some nontrivial theory assumptions, it is not totally
surprising to find different compositions of the wave functions, yet it is still frustrating. The best short- or
midterm prospects for improvement lie in obtaining constraints from more data of even greater
variety.

 Let us now provide some estimates of how much
future data can reduce most of the uncertainties discussed here.

\subsection{Improving the Constraints of the $\eta-\eta^{\prime}$ Wave Functions}
\label{IMPROV}
The determination of mixing angles and gluonium content is based on measurements.
The significance of such constraints depends on the experimental uncertainties.
Therefore we analyze which experimental inputs will best improve our knowledge of the $\eta-\eta^{\prime}$
wave functions. We start with the PDG 2010 values \cite{PDG10}:
\begin{itemize}
\item
The stated $ \phi \to \eta^{\prime} \gamma $ partial width is mainly due to the KLOE
measurement in \cite{kloe07}; the error is dominated by systematics due to the secondary $\eta^{\prime}$
branching ratio. The $\phi \to \eta \gamma$ branching ratio has been accurately measured by CMD-2 and SND \cite{PDG10}.

\item
The $ \eta^{\prime} \to \omega \gamma $ partial width of $(0.0053\pm 0.0005)$ MeV
with a relative error of $9\%$ comes from the overall PDG 2010 fit.
The relevant experiment has been performed in 1977 and was based on 68 events
\cite{Zanfino:1977ji}.
The KLOE-2 Collaboration \cite{AmelinoKloe2} could measure the branching ratio
$ {\cal{B}} (\eta^{\prime} \to \omega \gamma)$ more accurately by collecting at least 20 fb$^{-1}$ of data;
the limiting factor then comes from the uncertainty in
the total $\eta^{\prime}$ width, $ \Gamma_{\eta^{\prime}}$, since it is the partial width that matters.

\item
The $ \eta^{\prime} \to \rho \gamma $ partial width inferred from the PDG 2010 fit is $(0.0568\pm 0.0030)$ MeV;
the absolute branching ratio measurement was performed in 1969 by Rittenberg
\cite{Ritt} based on 298 events.
The PDG fit value is slightly lower than the directly
measured one.
Again the error is dominated by the uncertainty in $\Gamma_{\eta^{\prime}}$.

\item
The latest values on $ \rho \to \eta \gamma $ and $ \omega \to \eta \gamma $ partial widths are obtained in \cite{Achasov:2007kw},
based on SND data on $e^+ e^- \to \eta \gamma$: their accuracy is quite comparable to that of the PDG  2010 fit values.
\end{itemize}

In Table \ref{dectab3} we sketch different experimental scenarios. Starting from the present status as given by
PDG 2010 we analyze the impact various conceivable improvements in the experimental constraints would have on
the determination of the mixing angle $\phi_P$ and the size of $Z^{2}_{\eta^{\prime}}$, the gluonic component in the
$\eta^{\prime}$ wave function.
We have chosen the radiative processes that are common to analyses \cite{kloe07,Rafel,Thomas} discussed in Sect.~\ref{PRESTAT}.

In column I we list the uncertainties in the experimental input values as stated in PDG 2010.
In column II we indicate the improvement
that could be achieved by studying $\eta^{\prime}\to\omega\gamma$ with a sample of 20 $\rm{fb}^{-1}$ of $e^+ e^- \to \phi$ events,
that KLOE-2 anticipates to acquire  in the next few years \cite{AmelinoKloe2}. We
assume a selection efficiency of order $20\%$ in the analysis of $\phi \to \eta^{\prime} \gamma $ with $\eta^{\prime} \to \omega
\gamma $ and neglect background subtraction.
We observe that the limiting factor  is provided  from the uncertainty in
the total $\eta^{\prime}$ width.
In column III we indicate the improvement
that could be achieved by reducing the uncertainty on $\eta^{\prime}\to\rho\gamma$
of one-half respect to the present scenario; such improvement is also possible after a few years of running of
KLOE-2  \cite{AmelinoKloe2}.
In column IV and V we indicate the sensitivity
to an improvement in the determination of the partial widths
for $\phi \to \eta^{(\prime)}\gamma$ and for all the partial widths, respectively.
Among possible secondary decays of $\phi \to \eta^{\prime}\gamma$, there are both decays $ \eta^{\prime} \to \rho \gamma $
and $ \eta^{\prime} \to \omega \gamma $, whose errors are  dominated by the uncertainty on $\Gamma_{\eta^\prime} $.
However, the former is more convenient to measure, e.g. at KLOE, since it has a branching ratio of almost an order of magnitude larger;
also the total $\rho$ decay width $\Gamma_\rho$ is much larger, partially including and obscuring, from an experimental point of view,
the total $\omega$ decay  width $\Gamma_\omega$.

Since the partial widths of processes containing $\eta^\prime$ and the total width $\Gamma_{\eta^{\prime}}$ are correlated,
in column VI we evaluate the impact of the reduction of the uncertainty on $\Gamma_{\eta^{\prime}}$.
We assume a future $\Gamma_{\eta^{\prime}}$ measurement with $1.4\%$ uncertainty, which is within the possibility of KLOE-2 \cite{Ambrosino:2009sc}.
Such a measurement allows a determination of a nonzero gluonium content at $5\sigma$, as shown in column VI.
The crucial quantities to consider are not the central values for $\phi_P$ and $Z^{2}_{\eta^{\prime}}$, since they are likely
to shift, but their uncertainties. We conclude it is most important to reduce the uncertainty in the partial width for
$\eta^{\prime} \to \rho \gamma$; i.e., one has to measure both ${\cal{B}}(\eta^{\prime}\to \rho \gamma)$ and $\Gamma_{\eta^{\prime}}$
more accurately.

\begin{table}[htb]
\centering
\vskip 0.1 in
\begin{tabular}{|l|cccccc|} \hline
 \footnotesize  Processes &     $(\delta \Gamma/\Gamma)_I$ & $(\delta \Gamma/\Gamma)_{II}$ & $(\delta \Gamma/\Gamma)_{III}$ & $(\delta \Gamma/\Gamma)_{IV}$
      & $(\delta \Gamma/\Gamma)_{V}$& $(\delta \Gamma/\Gamma)_{VI}$\\
\hline
 \footnotesize $\phi \to \eta^{\prime}\gamma$ & $3.5\%$ &$3.5\%$ & $3.5\%$ &$1.7\%$ & $1.7\%$ & $1\%$\\
 \footnotesize $\phi \to \eta\gamma$ & $2\%$&$2\%$ & $2\%$ &$1\%$& $1\%$ & $2\%$\\
 \footnotesize $\eta^{\prime}\to\omega\gamma$ &$9\%$ &$4.5\%$& $9\%$ &$9\%$& $4.5\%$ & $1.7\%$\\
 \footnotesize $\eta^{\prime}\to\rho\gamma$ & $5\%$&$5\%$&$2.5\%$&$5\%$& $2.5\%$ &$1.7\%$\\
 \footnotesize  $\rho\to\eta\gamma$ & $7\%$&$7\%$&$7\%$ &$7\%$& $3.4\%$ & $7\%$\\
 \footnotesize $\omega\to\eta\gamma$ &$9\%$ &$9\%$&$9\%$ &$9\%$& $4.5\%$ &$9\%$ \\\hline
 \footnotesize $\phi_P$& \footnotesize $(40.6\pm0.9)^{\circ}$& \footnotesize $(40.1^{+0.8}_{-1.0})^{\circ}$&\footnotesize $(40.7\pm0.7)^{\circ}$& \footnotesize $(40.6^{+0.5}_{-0.6})^{\circ}$& \footnotesize $(40.4\pm0.5)^{\circ}$& \footnotesize $(40.1 \pm 0.3)^{\circ}$\\
 \footnotesize $Z^{2}_{\eta^{\prime}}$&\footnotesize $(0.09\pm0.05)$&\footnotesize $(0.13\pm0.05)$& \footnotesize $(0.08\pm0.04)$&\footnotesize $(0.09\pm0.03)$& \footnotesize$(0.10\pm0.03)$ & \footnotesize $(0.13\pm0.02)$ \\
\hline
\end{tabular}
\caption{ \it I: widths from PDG 2010 fits; II:  errors on   $\eta^{\prime}\to\omega\gamma$ reduced; III: errors on   $\eta^{\prime}\to\rho\gamma$ reduced; IV: errors on $\phi \to \eta^{(\prime)} \gamma$ reduced; V: reducing the uncertainties for all the partial widths; VI all recalculated in the hypothesis of
$1.4\%$ for the $\eta^{\prime}$ full width.}
\label{dectab3}
\end{table}
The situation concerning the $\eta^{\prime}$ full width is somewhat curious at present: PDG 2010 lists as its best value
$\Gamma_{\eta^{\prime}}=(0.194\pm0.009)$ MeV -- with the error including a scale factor of 1.2 -- resulting from an overall fit.
Direct measurements from 1979 \cite{Binnie:1979tk} and 1996 \cite{Wurzinger:1996jb} on the other hand yield the average
$\Gamma_{\eta^{\prime}}=(0.30\pm0.09)$ MeV, which would lead to $\phi_P=(42.7^{+1.0}_{-1.7})^{\circ}$ and $Z^{2}_{\eta^{\prime}}=(0.00\pm0.13)$.
Recently a new measurement has been performed at the COSY-11 facility:
$\Gamma_{\eta^{\prime}}=0.226\pm0.017(stat)\pm0.014(syst)$ MeV; the value of the width was established directly
from the measurement of the mass distribution of the $\eta^{\prime}$ meson, determined with a very high resolution
\cite{COSY11}.
The present average world value (2011 PDG partial update) contains this last measurement and gives
$\Gamma_{\eta^{\prime}}=(0.199\pm 0.009)$;
in the  global fit to the $\eta^{\prime}$ partial widths the correlations among the partial widths do not change
significantly.

Let us observe that the total width $ \Gamma_{\eta^{\prime}}$ extracted by PDG and the value of the partial width
$\Gamma( \eta^\prime \rightarrow \gamma \gamma)$ are strongly correlated, which may create difficulties when the total
and the partial width are used at the same time, as in the present case of the mixing angle extraction.
Moreover, the branching ratios of the $\eta^\prime$ meson decay channels are generally known with a relative precision
of more than an order of magnitude better than the present accuracy with which $ \Gamma_{\eta^{\prime}}$ is extracted.

\section{Weak Decays of Charm and Beauty Hadrons}
\label{WEAKDBDEC}

After many years of strenuous efforts to obtain the $\eta$ and $\eta^{\prime}$  wave functions with
nontrivial bounds why should one
not declare ``victory'' and go on to something else? There are three reasons:
\begin{itemize}
\item
Professional pride -- not to be belittled in Italy and Bavaria.
\item
Lattice QCD simulations have just entered the adult period.
\item
Yet there is the most topical reason, namely that knowing reliably the
$\eta$ and $\eta^{\prime}$  wave functions are an important input for our
understanding of several {\em weak} decays of beauty and charm hadrons.
Most crucially we need it for predicting  CP asymmetries involving
$\eta$ and $\eta^{\prime}$ in the final states and to understand whether
a deviation from standard model (SM) predictions can be seen as a signal of physics beyond the SM
\cite{KouGerard,interest1,interest2,interest3}.

The SuperB and Super KEK B factories approved in Italy and in Japan, respectively, will produce crucial statistics needed
for $B_{(s)} \to \eta/\eta^{\prime}X$ and $D_{(s)} \to \eta/\eta^{\prime}X$. There is a good
chance that LHCb will likewise and much sooner.

\end{itemize}

\subsection{Semileptonic Modes}
\label{SLDEC}

Since one expects semileptonic transitions to be driven by SM dynamics only (or at least to a high
degree of accuracy), their detailed studies teach us lessons on how nonperturbative hadronization
transforms quark level transitions. We will analyze here what semileptonic $D$ and $B$ decays
can tell us about the $\eta$ and $\eta^{\prime}$ wave functions and maybe more importantly,
how our knowledge of those can help us to better understand the decay mechanisms.

Before going into a more detailed discussion, a few general points should be mentioned. The
transitions $D_s^+ \to \eta^{(\prime )} l^+ \nu$, $D^+\to \eta^{(\prime )} l^+ \nu$ and
$B^+ \to \eta^{(\prime )} l^+ \nu$ proceed on greatly different time scales, since they are driven by
weak interactions on the Cabibbo-allowed, Cabibbo-suppressed and Kobayashi-Maskawa-suppressed levels,
respectively. Yet they can provide us with highly complementary information in the sense that
they produce the $\eta^{(\prime )}$ via their $s \bar s$, $d \bar d$ and $u \bar u$ components, respectively.
In addition, as explained below, $\eta^{(\prime )}$ could be excited via a
$gg$ component.

\subsubsection{$D_{(s)} \to \eta^{(\prime)}l\nu$}
\label{SLCHARM}

\begin{figure}[t]
 \centering
 \includegraphics[scale=0.45]{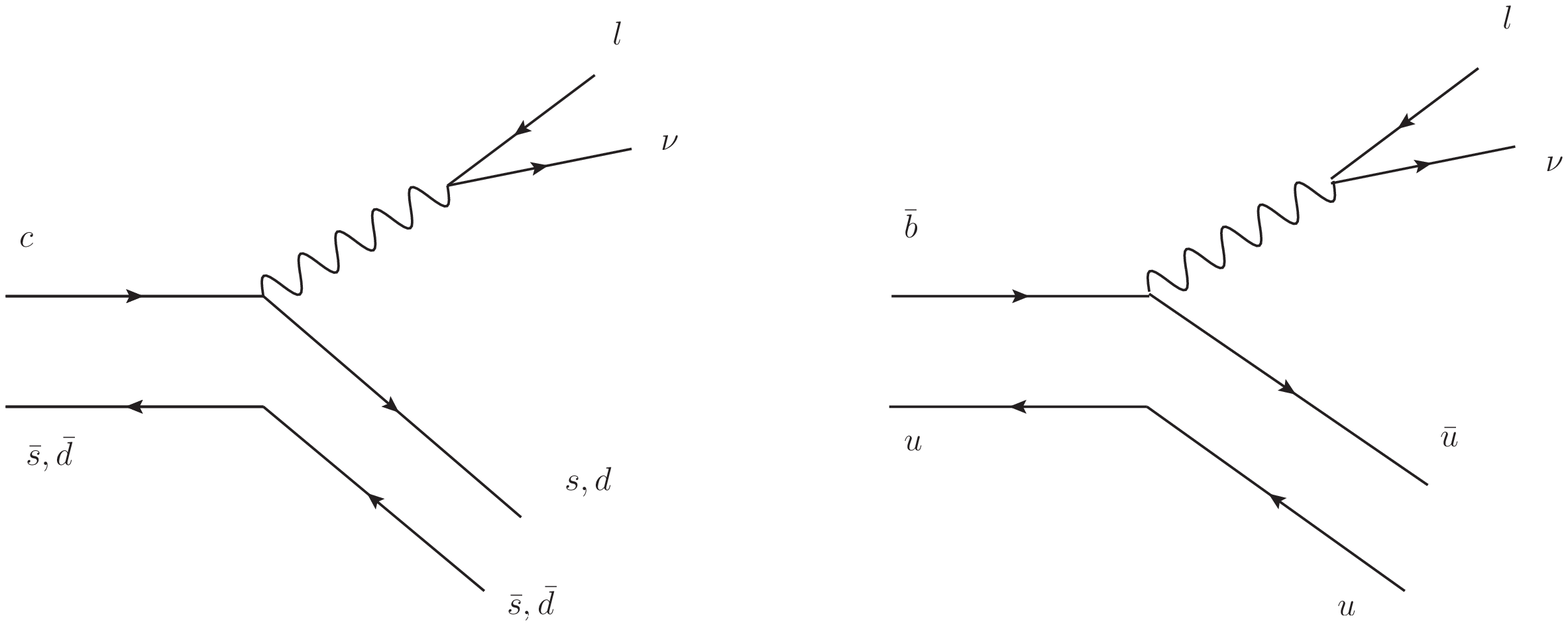}
\caption{\it Spectator diagrams for $D_{(s)} \to \eta^{(\prime)}l\nu$  and $B^+ \to \eta^{(\prime)}l\nu$  decays.}
\label{fig:diag:spectator}
\end{figure}

According to the heavy quark expansion the so-called spectator diagrams
(see e.g. Fig.\ref{fig:diag:spectator}) provide the leading
contribution to semileptonic  as well as nonleptonic charm decays \cite{cice}.

Data on semileptonic decays need to improve greatly before they
can constrain the physics related to the mixing with the gluonic component.
In 1995 CLEO extracted the branching fraction ${\cal{B}} ( D_s^+ \rightarrow \eta^{(\prime )} e^+ \nu)$
from ratios to hadronic decays of the $D^+_s$ \cite{cleo}. In 2009 CLEO-c presented the first absolute
measurement of the branching fraction of $ {\cal{B}} ( D_s^+ \rightarrow \eta^{(\prime )} e^+ \nu)$ \cite{:2009cm}; the  ratio
\beq
\left. \frac{ {\cal{B}}(D_s^+ \rightarrow  \eta^\prime e^+ \nu)}{{\cal{B}}(D_s^+ \rightarrow  \eta e^+ \nu)}
\right|_{\rm CLEO-c} = 0.36 \pm 0.14
\label{newCleodata}
\eeq
is in
agreement with the previous CLEO results \cite{cleo}.
In semileptonic $D_s$ decays
the final state hadron has to be produced off an $s \bar s$ configuration;
if $\eta$ and $\etp$ are pure $q\bar q$ states, i.e. $Z^2_{\eta^{\prime}} = Z^2_{\eta }= 0$,  then
one finds
in the quark flavor basis
\begin{equation}
\frac{\Gamma(D_s^+ \to \eta^{\prime} e^+ \nu)}{\Gamma(D_s^+ \to \eta e^+ \nu)} = R_D \cot ^2 \phi
\label{ratiowidth}
\end{equation}
with the quantity $R_D$ given by the relative phase space and the ratio of the $\eta$ and $\etp$
form factors integrated over the appropriate range in $q^2$.
To calculate the explicit form of $R_D$ one has to model the $q^2$ dependence of the form factors, but the factorization of the mixing angle dependence can help to devise tests of the mixing angle itself (see e.g. \cite{datta}). From the previous CLEO results \cite{cleo}, using $\eta$ and $\etp$ as pure $q\bar q$ states
and a pole ansatz for the form factors
Feldmann, Kroll and Stech inferred
$
\phi_P = (41.3 \pm 5.3)^{\circ}
$ \cite{Feld&S};
 it agrees even better than one might have expected with the values given above as extracted from weak and electromagnetic transitions.
Their value  is consistent with the new CLEO data within the errors.

Gronau and Rosner in a  very recent paper \cite{GROROS} gave a similar number for
$\Gamma(D_s^+  \to \eta^{\prime} l^+ \nu)/\Gamma(D_s^+  \to \eta l^+ \nu)$ (among other predictions) applying a very
simple model, where $R_D$ is inferred from kinematic factors in the quark level; again,
$\eta$ and $\etp$ are described as pure $q\bar q$ states.

The  transition form factors encode complex
hadronic dynamics and momentum dependence: in \cite{Anisovich:1997dz} they have been expressed through the
light-cone wave functions of the initial and final mesons. An allowed range for $Z_\eta^2/Z_{\eta^\prime}^2$ is given;
at the point $Z_\eta^2=0$, the angle $\phi_P$ is estimated to be $\phi_P = (37.7 \pm 2.6)^{\circ} $ and the simple factorized relation holds \cite{Anisovich:1997dz}
\beq
\frac{\Gamma(D_s^+  \to \eta^{\prime} e^+ \nu)}{\Gamma(D_s^+  \to \eta e^+ \nu)} = R_D \cot ^2 \phi_P \cos^2\phi_G
\label{withgluon}
\eeq
where $\phi_G$ has been defined in Eq. (\ref{colla1}). In \cite{Anisovich:1997dz} the value
 $R_D=0.28$ is estimated by neglecting the nontrivial dependence on
the constituent quark transition form factor, that is a conventional approximation in literature, while
 $R_D= 0.23$ is estimated by assuming a simple monopole $q^2$ dependence.
We observe that the mixing angle extracted from (\ref{withgluon}) is strongly dependent on the value of $R_D$; in order to provide a rough estimation of the theoretical error we consider an averaged $R_D$, that is $R_D=0.255 \pm 0.050$.
By using the experimental ratio of branching fractions (\ref{newCleodata}), we estimate
$Z^{2}_{\eta^{\prime}}=0.16\pm0.33_{exp}\pm 0.23_{th}$, that is
  $\phi_G = (23.3 \pm 25.8_{exp}\pm 18.0_{th})^{\circ}$, where  the theoretical error refers to the errors on $R_D$ and $\phi_P$ added  in quadrature.
The experimental error dominates over the rough estimate of the theoretical error and it prevents any conclusion on the gluonic content of the $\eta-\etp$ system.


For the Cabibbo suppressed transitions one finds in the same framework:
\begin{equation}
\frac{\Gamma(D^+ \to \eta^{\prime} e^+ \nu)}{\Gamma(D^+ \to \eta e^+ \nu)} =   {\tilde{R}}_D  \tan ^2 \phi_P
\label{ratiowidth2}
\end{equation}

In 2008 CLEO-c reported its first measurement of
$\Gamma(D^+ \to \eta e^+ \nu)$ and an upper bound on $\Gamma(D^+ \to \etp e^+ \nu)$
\cite{cleoc}.
Two years later, the same collaboration presented
the first observation of $ D^+ \rightarrow \eta^\prime e^+ \nu$,
with  branching fraction ${\cal{B}}  (D^+ \to \etp e^+ \nu) = (2.16 \pm 0.53 \pm 0.07) $ x $  10^{-4}$,
and an improved ${\cal{B}} (D^+ \rightarrow \eta e^+ \nu) =
(11.4 \pm 0.9 \pm 0.4)$ x $ 10^{-4}$ \cite{:2010js}.
By using the above data and the reasonable assumption  $R_D\simeq \tilde R_D$, we estimate from Eq. (\ref{ratiowidth2}) the value $\phi_P = (41 \pm 4_{exp} \pm 3_{th})^{\circ}$.

 By including a nonzero gluon contribution, we can parametrize the $D^+$ ratio as in (\ref{withgluon}).
However, with the available recent data, the estimate of the angle $\phi_P$ can be made independently of $\phi_G$
 by taking the ratio
 \beq
 \frac{\Gamma(D_s^+  \to \eta^{\prime} e^+ \nu)/\Gamma(D_s^+  \to \eta e^+ \nu)}{\Gamma(D^+ \to \eta^{\prime} e^+ \nu)/\Gamma(D^+ \to \eta e^+ \nu)} \simeq \cot^4 \phi_P
 \eeq
The left side is given by the recent experimental data quoted before, and we get $\phi_P = (40 \pm 3)^\circ$.

Yet this is not the final word on the experimental or theoretical side.
A few years down the line we can expect BESIII to obtain an even larger sample allowing a more
accurate measurement with errors on the angle $\phi_P$ going down to about $2 \%$.

\begin{figure}[t]
 \centering
 \includegraphics[scale=0.85]{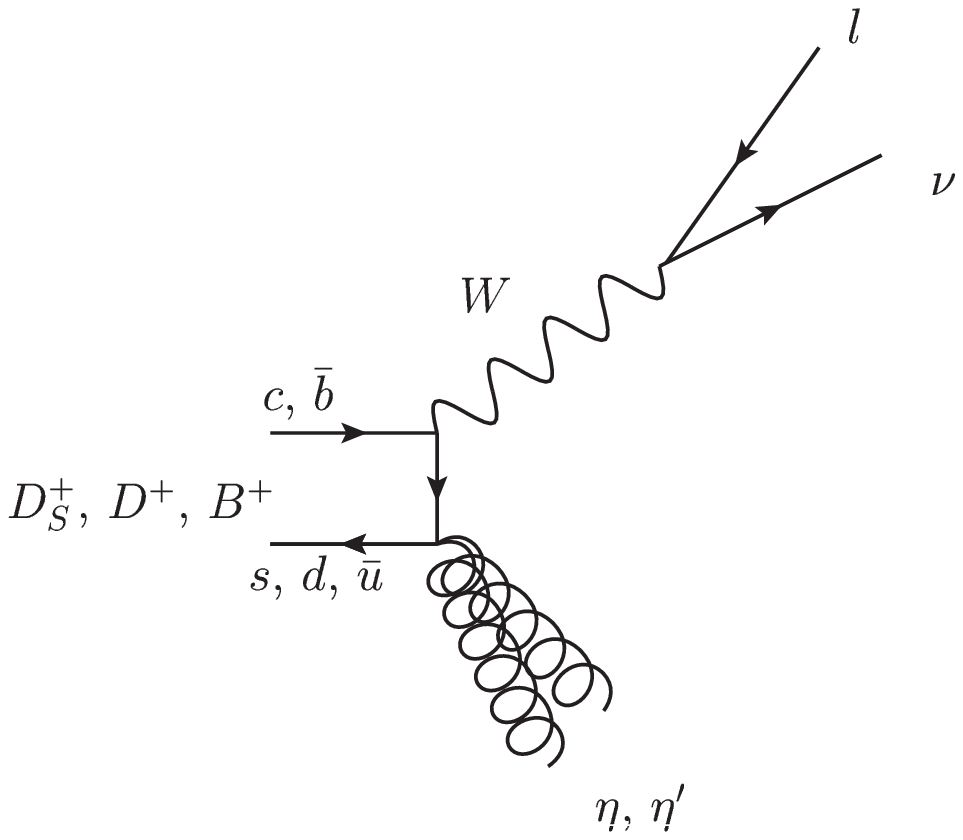}
\caption{\it Valence quarks $c/\bar s/\bar d$ (as well as $\bar b/u$)
        emitting two gluons which generate $\eta/\eta^{\prime}$ via the gluonic component of the wave functions.}
\label{fig:diag}
\end{figure}

The theoretical
situation is more complex. While the spectator diagram generates the leading contribution,
for a precision study we cannot ignore non-leading ones. The so-called ``weak annihilation" (WA) process contributes even to semileptonic meson decays \cite{wapaper,cice}, as can be illustrated
most directly for $D_s^+$ and $D_s$; see Fig. \ref{fig:diag}.
An analysis based on inclusive semileptonic $D$ decays, which considers
both the widths and the lepton energy moments, shows no clear evidence of WA effects \cite{Gambino:2010jz}. While WA might affect the corresponding
inclusive semileptonic width only moderately, it should impact the exclusive channels
$D_s^+ \to \etp l^+\nu $ and  $D^+ \to  \etp l^+\nu $ on
the Cabibbo-favoured and suppressed levels via the $\etp$'s gluonic component. The strength of the
effect depends on two factors, namely, the size of the $gg$ component in the $\eta^{\prime}$
wave function and on how much $gg$ radiation one can expect in semileptonic $D_s^+$, $D^+$ and
$B^+$ decays. Lastly, since the main effect might come from the interference with the spectator amplitude, it
can a priori enhance or reduce those rates. Simple relations such as  (\ref{ratiowidth}) do not necessarily hold any longer.

\subsubsection{$B^+ \to \eta^{(\prime)}l\nu$}
\label{SLBEAUTY}

In $B^+\to \eta^{(\prime)}l\nu$ decays one encounters a situation analogous to that for $D^+\to \eta^{(\prime)}l\nu$
except that their rates are suppressed by $|V_{ub}/V_{cb}|^2$ rather than $|V_{cd}/V_{cs}|^2$ and that the range
in $q^2$ is much larger. In passing we just want to mention that one needs to understand their rates to determine
$|V_{ub}/V_{cb}|^2$ from $\Gamma (B \to X_u l \nu )/\Gamma (B \to  X_c l \nu )$ with the hoped-for accuracy
of about 5\% \cite{GAMB}.

In the spectator ansatz one finds using the quark-flavor basis
\begin{equation}
\frac{\Gamma(B^+ \to \eta^{\prime} l^+ \nu)}{\Gamma(B^+ \to \eta l^+ \nu)} = \tilde R_B \tan ^2 \phi
\label{ratiowidth3}
\end{equation}
with the factor $\tilde R_B$ again describing the relative phase space (much more abundant than for
$D$ mesons) and the ratio of the integrated form factors.
The semi-leptonic form factors $ B \rightarrow \eta^{(\prime)}$ have been calculated in
\cite{Ball:2007hb} from QCD sum rules
on the light-cone, to next-to-leading in QCD. In frameworks based on QCD factorization the mesons
Fock-state wave functions enter in the form of light-cone distribution amplitudes. Equation (\ref{ratiowidth3}) keeps robust under the dynamical assumptions in \cite{Ball:2007hb}.
Data on the ratio (\ref{ratiowidth3}) have started to appear since a few years. The errors are still quite large, comparable in percentage to the ones analyzed in the previous section, and prevent definite conclusions on the glue mixing to be drawn.

In 2007 CLEO has found first evidence for $B^+ \rightarrow  \eta^\prime l^+ \nu$ decay,
with branching fraction ${\cal B}(B^+ \rightarrow  \eta^\prime l^+ \nu) = (2.66 \pm 0.80 \pm 0.56) \times 10^{-4}$.
This year, also the
BABAR Collaboration measured for the first time
 ${\cal B}(B^{+} \rightarrow \eta^{\prime} l^{+} \nu) = \left(0.24 \pm 0.08_{stat} \pm 0.03_{syst} \right) \times 10^{-4}$ \cite{delAmoSanchez:2010zd},
superseding the 2008 upper limit \cite{:2008gka}.
The BABAR value has a significance of $3.0 \sigma $ and
 it is an order
of magnitude smaller than the CLEO result.

\noindent
The same 2007 CLEO analysis also reported a new value of the branching fraction
${\cal B}(B^+ \rightarrow  \eta l^+ \nu) = (0.44 \pm 0.23 \pm 0.11) \times 10^{-4}$ \cite{Adam:2007pv}, improving previous 2003 values \cite{Athar:2003yg}. The result is similar to the newest one by BABAR:
${\cal B}(B^{+} \rightarrow \eta l^{+} \nu) = \left(0.36 \pm 0.05_{stat} \pm 0.04_{syst} \right) \times 10^{-4}$ \cite{delAmoSanchez:2010zd}.
By using BABAR data  \cite{delAmoSanchez:2010zd}, the ratio (\ref{ratiowidth3}) reads \beq
\left. \frac{ {\cal{B}}(B^+ \to \eta^{\prime} l^+ \nu)}{{\cal{B}}(B^+ \to \eta l^+ \nu)}
\right|_{\rm BaBar}
 = 0.67 \pm 0.24_{stat} \pm 0.11_{syst}  \eeq
 It is evident that the experimental situation is not yet satisfying, although the previous value does not exclude
 a large gluonic singlet
contribution to the $\eta^\prime$ form factor.

The corresponding ratio involving the $B_s$ mesons, that is
$ {\cal B}(B_s \rightarrow  \eta^\prime l^+ l^-)/{\cal B}(B_s \rightarrow  \eta l^+ l^-) $, is also
potentially informative on the $\eta^{(\prime)}$ gluonic content, although
 experimentally much more challenging.
The results for the branching fractions of modes with two
charged leptons in the standard model are of order $10^{-7}-10^{-8}$  \cite{Carlucci:2009gr},
 suggesting that they are within the reach of SuperB and Super KEK B factories.

\subsection{Non-leptonic $D$ and Charmless $B$ Decays}
\label{NONLEP}

Although estimates of the mixing angles may come from $ b \rightarrow c$ dominated processes, such as
 $B^0_{s} \to J/\psi \eta^{(\prime)}$ (see e.g. \cite{Thomas, datta, newFKR}),
within the SM many charmless nonleptonic $B$ decays receive significant or even leading contributions
from loop processes, which represent
quantum corrections. Thus they provide fertile hunting grounds for new physics, in particular in their
\cp~asymmetries. Yet to make a convincing case that an observed \cp~asymmetry is such that it could
not be generated by SM forces alone, one has to be able to evaluate hadronic matrix elements.
Such an undertaking is greatly helped by knowing the wave functions of the relevant particles.

\begin{figure}[t]
 \centering
 \includegraphics[scale=0.65]{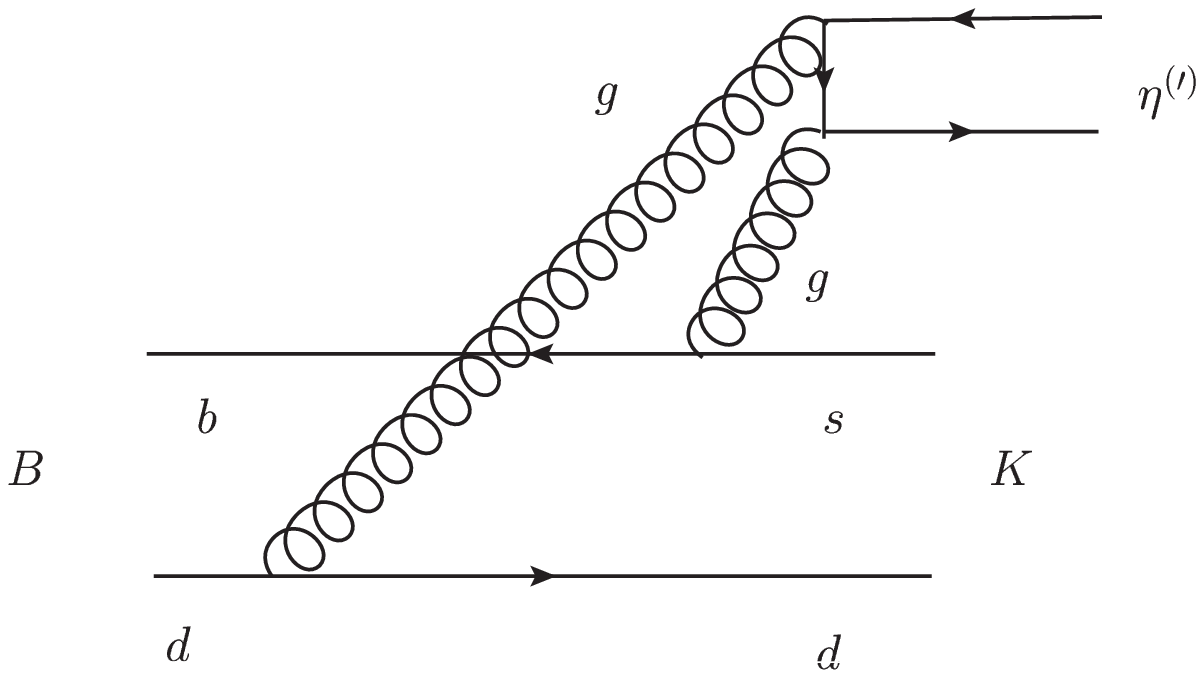}
   \caption{\it $\eta^{(\prime)}$ produced via its gluonic component with the gluons being radiated
        off different quark lines. The transition $b \rightarrow s$ is penguin mediated. }
\label{fig:ggfusion}
\end{figure}

Modes such as $B \to \eta^{(\prime)}K$ and $B_s \to \eta^{(\prime)}\phi$ seem particularly well suited in this
respect.
It should be noted that the branching ratio observed for $B \to \eta^{\prime}K$ exceeds
the original predictions considerably for reasons that have not been established yet. Those predictions
had been based (among other assumptions) on identifying $\etp$ as a pure $q\bar q$ state.
Allowing for a gluonic component opens the door to diverse decay mechanisms. For example,
Kou and Sanda \cite{K&S} suggested producing the $\eta^{\prime}$ meson via its gluonic component with the gluons being radiated off {\em different} quark lines, see Fig. \ref{fig:ggfusion}.
Having the gluon radiation being emitted from a single quark line might be a more favorable dynamical
scenario (see e.g. \cite{Dighe:1995gq,Gronau:1995ng}).


%
%
%

Recent branching ratios values are ${\cal B}(B^0\rightarrow K^0 \eta) = \left(1.1 \pm 0.4 \right) \times 10^{-6}$ and ${\cal B}(B^0\rightarrow K^0 \eta^\prime) = \left(6.6 \pm 0.4 \right) \times 10^{-5}$ \cite{PDG10}.
 The $B \to \eta^{(\prime)}K$ decays may proceed through tree diagrams $ \bar b \rightarrow \bar u u \bar s$, but such contributions are colour and Cabibbo-Kobayashi-Maskawa suppressed, and by one-loop $ b \rightarrow s$ penguins. Although the same basic penguin mechanism is expected to drive both $B \to \eta^{(\prime)}K$ and $B\to \pi K $, the rate of the former is measured to be much larger. A possible distinctive contribution are flavor singlet amplitudes that are not allowed, if the final state contains only flavor nonsinglet states such as pions and kaons.
In flavor singlet penguins two gluons couple to the $\eta^\prime$ violating the OZI rule and the amplitude can get contributions from the pure gluonic component of the $\eta^\prime$.

The cases where two gluons are emitted by a single line ($ b \rightarrow s \, g \, g$) together with spectator scattering and singlet weak annihilation have been explored in the context of QCD factorization (QCDF) \cite{Beneke:2002jn}. In this approach the constructive interference between nonflavor singlet penguins seems already sufficient to enhance the
$B \to \eta^{\prime}K$ branching ratio, without the recourse to flavor singlet contributions; however, due to large hadronic uncertainties, a sizable gluonic contribution (up to $40 \%$) to the $ B \rightarrow \eta^\prime$ form factor cannot be excluded.

In the perturbative QCD approach the impact of the gluonic component
on the branching ratio - potentially important since it
 increases the branching ratios $B \to \eta^{\prime}K$, while decreasing the $\bet$ one - has been estimated to be numerically very small \cite{Charng:2006zj}.
The phenomenological importance of the $\eta^\prime$ gluonic
content was instead emphasized in the context of soft collinear effective theory (SCET) \cite{Williamson:2006hb}.

Let us note that the previous exclusive analyses have been performed not later than 2006, when relevant new data, such as semileptonic $ B \rightarrow \eta \, l \, \nu$ branching ratios, were not yet available. In semileptonic decays there is no enhancement in the $B$ decay into $\eta^\prime$ mesons. The enhancement is also not observed in $ D_s^+ \rightarrow K^+ \eta^\prime$ relative to
$ D_s^+ \rightarrow K^+ \pi^0$.  Recent data from BABAR for decays into
$K^\star$ \cite{delAmoSanchez:2010qa}
favor an opposite pattern with respect to $K$, namely
$\Gamma( B \rightarrow K^\star \eta^\prime) < \Gamma( B \rightarrow K^\star \eta)$.
It  would be interesting to check the impact of all recent experimental values on the different approaches. For instance in
\cite{Williamson:2006hb}, the effort to fix the size for the gluonic contribution to the $ B \rightarrow \eta^\prime$ form factor, in a more constrained way with respect to \cite{Beneke:2002jn},
partly depends on fitting nonperturbative parameters to experimental data. We have to admit that a quantitative comparison with data is hampered by the theoretical uncertainties in nonleptonic decays.

The large measured branching ratio for $B^0\rightarrow K_S \eta^\prime$ by the BABAR and Belle Collaborations \cite{:2008se, Chen:2006nk} greatly improves the usefulness of the decay mode for measuring CP asymmetry and to produce significant deviation from the SM prediction.
The projected SuperB and Super KEK B factories will probe highly nontrivial ranges for new dynamics.

While it is true that the size of the time-dependent \cp~asymmetry established in
$B_d \to \eta^{\prime}K_S$ conforms well with the SM expectation, one cannot count on an intervention
of new physics being numerically large there. Having a smallish deviation being significant implies good theoretical control over the SM prediction,  which in turn requires good knowledge of the $\etp$ as well as $\eta$ wave functions.

Finding \cp~asymmetries in $D\to \eta^{(\prime)}\pi$, $\eta^{(\prime)}\eta$, $\eta \phi$ and interpreting them as
signals of new dynamics has two experimental and theoretical advantages:
\begin{itemize}
\item
The branching ratios are not very small.
\item
The SM can produce only very tiny \cp~asymmetries. Even small asymmetries produce clear signatures for
new physics, as long one can control systematic uncertainties.
\end{itemize}

\section{Summary}
\label{OUT}
In this paper we have described the status of ongoing investigations, starting from a review of the
knowledge on the $\eta$ and $\eta^{\prime}$ wavefunctions in terms of quark and gluon
components as has been inferred mainly from radiative $\phi$ and $\psi$ decays.
The different determinations of the $\eta-\eta^{\prime}$ mixing are generally consistent,
but the message concerning the gluon content in the $\eta^{\prime}$ remains ambivalent.
The semileptonic $D^+$, $D^+_S$ and $B^+$ decays can give other constraints to check
$\eta^{\prime}$
gluonium role. Moreover a sizable gluonium content could help to understand the unexpected high value
of the branching ratio for $B \to \eta^{\prime} K_S$ decay.

In conclusion: after many and difficult efforts to understand the $\eta-\eta^{\prime}$ wave functions it might be
seen as ``smart" to call it a ``victory" and move to another problem. We want to emphasize that it is a
``noble" goal to improve
our understanding of non-perturbative effects in QCD, in particular when more ``allies" from lattice QCD come
to the battle line. Furthermore, and maybe even more important, it will help significantly to identify
the footprints of new physics in CP asymmetries in $B$ and $D$ decays.

\section{Acknowledgements}

This work was supported by the NSF under the Grant No. PHY-0807959.
C.D.D. thanks Professor M. Wayne, Director of the
 Department of Physics, for the warm hospitality during her stay at the University of Notre Dame du Lac.
 We also thank Jonathan Rosner for
comments on the manuscript.

%
\end{document}